\documentstyle[preprint,aps]{revtex}
\tighten
\draft
\widetext
\preprint{CLNS 98/1563, HUTP-98/A049, NUB 3181}
\bigskip
\bigskip
\begin{document}
\title{Three Generations in Type I Compactifications}
\medskip
\author{Zurab Kakushadze$^{1,2}$\footnote{E-mail: 
zurab@string.harvard.edu} and S.-H. Henry Tye$^{3}$\footnote{E-mail: 
tye@mail.lns.cornell.edu}}
\bigskip
\address{$^1$Lyman Laboratory of Physics, Harvard University, Cambridge, 
MA 02138\\
$^2$Department of Physics, Northeastern University, Boston, MA 02115\\
$^3$Newman Laboratory of Nuclear Studies, Cornell University, Ithaca, NY 14853}
\date{June 17, 1998}
\bigskip
\medskip
\maketitle

\begin{abstract}
{}Generalizing the recent work on three-family Type I compactifications, we classify perturbative Type I vacua obtained via compactifying on the $T^6/{\bf Z}_2\otimes {\bf Z}_2\otimes {\bf Z}_3$ orbifold with all possible Wilson lines. In particular, we concentrate on models with gauge groups containing the Standard Model gauge group $SU(3)_c\otimes SU(2)_w\otimes U(1)_Y$ as a subgroup. All of the vacua we obtain contain D5-branes and are non-perturbative from the heterotic viewpoint. The models we discuss have three-chiral families. We study some of their phenomenological properties, and point out non-trivial problems arising in these models
in the phenomenological context.
\end{abstract}

\section{Introduction}

{}Recent progress in understanding Type IIB orientifolds \cite{group} in six \cite{PS} and four \cite{BL,Sagnotti,ZK,KS1,Zw,Iba,KST,large,su6,class,4-2-1} dimensions has shed a lot of light on ${\cal N}=1$ supersymmetric Type I compactifications in four dimensions. These developments have important implications for understanding non-perturbative string dynamics which is due to the
conjectured duality between Type I and heterotic superstrings \cite{PW}. One of the exciting consequences of this duality is that some non-perturbative heterotic vacua can be described via perturbative Type I compactifications. Thus, heterotic NS 5-branes which have no world-sheet description are dual to Type I D5-branes which are perturbative objects from the Type I viewpoint.

{}It is natural to expect that these developments might also have important implications for
phenomenological applications of string theory. Thus, phenomenologically oriented model building in superstring theory for many years had been mostly confined to the perturbative heterotic superstring framework where it was facilitated by the existence of relatively simple rules. In particular, perturbative heterotic superstring enjoys the constraints of conformal field theory and modular invariance which serve as guiding principles for model building. Given the success of string dualities in shedding light on non-perturbative string dynamics, it is natural to attempt construction of phenomenologically interesting superstring vacua which would be non-perturbative from the heterotic viewpoint. This, in particular, could {\em a priori} provide us with clues for solving some of the phenomenological problems encountered in heterotic model building. 

{}Phenomenological properties of Type I vacua are expected to be rather different than those of perturbative heterotic superstring. In particular, in Type I models, gravity lives in a ten dimensional bulk, whereas the gauge theories live on D9- as well as D5-branes. The key difference between perturbative heterotic superstring, where both gravity and gauge theory are ten dimensional, and Type I superstring is then that the observable gauge group may live in the world-volumes of some of the D5-branes. The implications of this fact are rather dramatic. In particular, the patterns of gauge and gravitational coupling unification in Type I are rather different from those in the perturbative heterotic superstring \cite{witten}. (For recent developments of this idea within both field and string theory approaches, see, {\em e.g.}, \cite{TeV,su6,ST,4-2-1,lykken}.)

{}It is reasonable to attempt construction of ``semi-realistic'' Type I vacua with at least some phenomenologically appealing features. One of the most desirable and basic phenomenological properties one would like to obtain is an acceptable gauge group and number of chiral families. The first step in this direction was made in \cite{su6} where a Type I model with an $SU(6)$ gauge subgroup and three-families of chiral fermions was constructed. The geometry of the corresponding compactification was that of the $T^6/{\bf Z}_2\otimes {\bf Z}_2\otimes {\bf Z}_3$ orbifold. 
This model contains D5-branes and is non-perturbative from the heterotic viewpoint.
The construction in \cite{su6} assumed a background with trivial Wilson lines. Recently, an example of Type I on $T^6/{\bf Z}_2\otimes {\bf Z}_2\otimes {\bf Z}_3$ with a non-trivial Wilson line was constructed in \cite{4-2-1}. This model contains an $SU(4)_c\otimes SU(2)_w \otimes U(1)$ gauge subgroup which is one of the phenomenologically acceptable gauge groups as it contains the Standard Model gauge group $SU(3)_c\otimes SU(2)_w \otimes U(1)_Y$. The number of families in the model of \cite{4-2-1} is three just as in its mother $SU(6)$ model of \cite{su6}. However, without a detailed phenomenological analysis it is hard to tell whether such Type I vacua have a chance to be realistic.

{}In this paper, which is a follow-up paper of \cite{4-2-1}, we systematically classify all 
perturbative Type I compactifications on $T^6/{\bf Z}_2\otimes {\bf Z}_2\otimes {\bf Z}_3$ that contain $SU(3)_c\otimes SU(2)_w \otimes U(1)_Y$. In particular, we study different possible types of Wilson lines that can be included in this compactification. We find about a dozen distinct vacua. We then investigate some of the phenomenological properties of these models.
In particular, we study the question of whether the gauge symmetry can be broken to  $SU(3)_c\otimes SU(2)_w \otimes U(1)_Y$, and also whether the ``naive'' three families present in the model indeed have the correct quantum numbers to be describing quarks and leptons. We also address the issue of extra ``vector-like'' matter whose decoupling turns out to be a rather non-trivial issue. In particular, we discuss some generic properties of these vacua. We also address the issue of baryon number violating terms which are generically expected to arise and would need to be suppressed in a realistic model. We point out that supersymmetry breaking in such vacua would require some rather non-trivial mechanisms. Our preliminary analyses here bring about many interesting issues that would need to be understood in the context of phenomenologically oriented Type I model building. 

{}The remainder of this paper is organized as follows. In section II we give an explicit construction of perturbative Type I on $T^6/{\bf Z}_2\otimes {\bf Z}_2\otimes {\bf Z}_3$ with all possible Wilson lines. In section III we discuss various phenomenological properties of these models. In section IV we summarize the main conclusions of this paper.

\section{Explicit Construction}

{}In this section we explicitly construct Type I vacua corresponding to the compactification on the ${\bf Z}_2\otimes {\bf Z}_2\otimes {\bf Z}_3$ orbifold. We will first review the case without  Wilson lines originally constructed in \cite{su6}. We then classify all the discrete Wilson lines compatible with the orbifold action, and construct the corresponding Type I vacua.

\subsection{Type I on the ${\bf Z}_2\otimes {\bf Z}_2\otimes {\bf Z}_3$ Orbifold.}

{}Consider a Type I compactification on the toroidal orbifold ${\cal M}=T^6/\Gamma$ with zero NS-NS $B$-field (that is, the internal components of the $B$-field are all zero: $B_{ij}=0$, $i,j=1,\dots,6$), and no non-trivial Wilson lines. Here $\Gamma\approx{\bf Z}_2\otimes 
{\bf Z}_2\otimes {\bf Z}_3$ is the orbifold group. Let $g$, $R_1$ and $R_2$ be the generators of the ${\bf Z}_3$ and the two ${\bf Z}_2$ subgroups of $\Gamma$. Then the action of $g$ and $R_s$
($R_3=R_1R_2$) on the complex coordinates $z_s$ is given by (here for the sake of simplicity we can assume that the six-torus $T^6$ factorizes into a product $T^6=T^2\otimes T^2\otimes T^2$ of three two-tori, and the complex coordinates $z_s$ ($s=1,2,3$) parametrize these three two-tori): 
\begin{equation}
 g z_s=\omega z_s~,~~~R_s z_{s^\prime}=-(-1)^{\delta_{ss^\prime}} z_{s^\prime}~.
\end{equation} 
(There is no summation over the repeated indices here.) Here $\omega=\exp(2\pi i/3)$. The Calabi-Yau three-fold ${\cal M}=T^6/\Gamma$ (whose holonomy is $SU(3)$) has 
Hodge numbers $(h^{1,1},h^{2,1})=(36,0)$. Thus, the closed string sector of Type I on ${\cal M}$ consists of ${\cal N}=1$ supergravity, the dilaton plus axion supermultiplet, and 36 chiral supermultiplets neutral under the open string sector gauge group\footnote{Some of these supermultiplets transform non-trivially under the anomalous $U(1)$'s in the open string sector, however.}.

{}The Type I model we are discussing here can be viewed as the $\Omega$ orientifold of Type II
B compactified on ${\cal M}$, where $\Omega$ is the world-sheet parity reversal. The tadpole cancellation conditions require introducing 32 D9-branes (corresponding to the $\Omega$ element of the orientifold group) as well as three sets of D5-branes (which we will refer to as D$5_s$-branes) with 32 D5-branes in each set. Moreover, the action of the orbifold group $\Gamma$ on the Chan-Paton charges carried by the D9- and D$5_s$-branes is described by
the corresponding Chan-Paton matrices $\gamma_{g_a}$ where $g_a$ are the elements of the
orbifold group $\Gamma$. These matrices must form a projective representation of (the double cover of) the orbifold group $\Gamma$. The twisted tadpole cancellation conditions give constraints on traces of the Chan-Paton matrices. In particular, we have the following tadpole cancellation conditions (here we note that the orientifold projection $\Omega$ on the D9-brane Chan-Paton charges is required to be of the $SO$ type in this model) \cite{su6}:
\begin{equation}
 {\mbox{Tr}}(\gamma_g)=-2~,~~~
 {\mbox{Tr}}(\gamma_{R_s})={\mbox{Tr}}(\gamma_{gR_s})=0~.
\end{equation}   
Here we are using the convention (which amounts to not counting orientifold images of D-branes) where the Chan-Paton matrices are $16\times 16$ matrices (and not $32\times 32$ matrices which would be the case if we counted both the D-branes and their orientifold images).
The solution (up to equivalent representations) to the twisted tadpole cancellation conditions is given by \cite{su6}:
\begin{equation}
 \gamma_{g,9}={\mbox{diag}}({\bf W}\otimes {\bf I}_3, {\bf I}_4)~,~~~
 \gamma_{R_s,9}=i\sigma_s \otimes {\bf I}_8~.
\end{equation}
Here ${\bf W}={\mbox{diag}}(\omega,\omega,\omega^{-1},\omega^{-1})$, $\sigma_s$ are the Pauli matrices, and ${\bf I}_n$ is an $n\times n$ identity matrix.
(The action on the D$5_s$-branes is similar.) The massless spectrum of this model was worked out in \cite{su6}. The gauge group is $[U(6)\otimes Sp(4)]_{99}\otimes \bigotimes_{s=1}^3 [U(6)\otimes Sp(4)]_{5_s 5_s}$. (Here we are using the convention where $Sp(2N)$ has rank $N$.) There are three chiral generations in each $SU(6)$ subgroup in this model.

\subsection{Cases with One Wilson Line}

{}Next, let us discuss turning on one discrete Wilson line in the above model. Thus, consider a freely acting orbifold which amounts to a ${\bf Z}_3$ shift in the
third $T^2$ (parametrized by the complex coordinate $z_3$). That is, let this two-torus be defined by the identifications $z_3\sim z_3+n^\alpha e_\alpha$, where $n^\alpha\in {\bf Z}$, and $e_\alpha$ ($\alpha=1,2$) are the constant vielbeins ({\em i.e.}, equivalent to radius times the $SU(3)$ simple roots). Then this ${\bf Z}_3$ shift $S$ has the following action on $T^2$: $Sz_3=z_3+{1\over 3} m^\alpha e_\alpha$
for some integers $m^\alpha$ such that ${1\over 3} m^\alpha e_\alpha\not\in \Lambda$, where
$\Lambda=\{n^\alpha e_\alpha\}$ is the lattice defining this torus (that is, $T^2={\bf C}/\Lambda$). This shift commutes with the ${\bf Z}_3$ twist generated by the orbifold group
element $g$, and also with the ${\bf Z}_2$ twist generated by the orbifold group
element $R_3$. However, it does not commute with the elements $R_1$ and $R_2$. In particular, $S$ and $R_1$ generate a non-Abelian group isomorphic to $D_3$. Similarly, 
$S$ and $R_2$ also generate a non-Abelian group isomorphic to $D_3$.

{}We need to specify the action of $S$ on the Chan-Paton charges.  
This action is subject to the corresponding tadpole cancellation conditions.
There is no restriction on the trace of $\gamma_S$ itself. (This can be seen, for instance,  by noting that in the $S$ and $S^{-1}$ ``twisted'' closed string sectors there are no massless states.)
However, there are constraints on the traces of $\gamma_{S^k g^{k^\prime}}$ ($k,k^\prime=1,2$):
\begin{equation}
 {\mbox{Tr}}(\gamma_{S^k  
 g^{k^\prime},9})={\mbox{Tr}}(\gamma_{g,9})={\mbox{Tr}}(\gamma_{g^{-1},9})=-2~.
\end{equation}  
(Similar constraints apply to the D$5_s$-brane Chan-Paton matrices as well.) The most general 
solution to these constraints reads (up to equivalent representations):
\begin{equation}
 \gamma_{S,p_s}={\mbox{diag}}({\bf W}^\prime\otimes {\bf I}_{2N_s}, 
 {\bf I}_{12-4N_s}, {\bf W}^\prime \otimes {\bf I}_{N_s}, {\bf I}_{4-2N_s})~,
\end{equation}
where ${\bf W}^\prime ={\mbox{diag}}(\omega,\omega^{-1})$. 
Here we have introduced  the following compact notation: $p_0=9$, and $s=0$ refers to the D9-branes; $p_s=5_s$ for $s=1,2,3$ so that these values of $s$ label the D$5_s$-branes.
Also, $N_s$ ($s=0,1,2,3$) can take values $0,1,2$. Note, however, that $N_s$ can be different for different $s$. In other words, the action of the Wilson line on different kinds of D-branes can be different. For example, we can have $N_0=0$ for the D9-branes, $N_1=1$ for the D$5_1$-branes, and $N_2=N_3=2$ for the D$5_2$- and D$_3$-branes. We will refer to these models as the $M_{N_0,N_1,N_2,N_3}$ models.

{}It is not difficult to see that the gauge group in a given $M_{N_0,N_1,N_2,N_3}$ model is 
$\bigotimes_{s=0}^3 [U(M_s)\otimes U(N_s)^3\otimes Sp(L_s)]_{ss}$, where
$M_s=6-2N_s$, and $L_s=4-2N_s$. Here the subscript ``$ss$'' refers to the fact that the corresponding factor comes from the D$p_s$-D$p_s$ open string sector. The massless open string spectra of the $M_{N_0,N_1,N_2,N_3}$ models are given in Table I in a compact 
notation\footnote{Our compact notation should be self-explanatory. However, the reader might benefit from viewing some examples in more explicit notations. Thus, in \cite{su6,4-2-1} two such models can be found. In particular, the $U(1)$ charges are explicitly displayed in those references.}. Note that these spectra are free of non-Abelian gauge anomalies. 
Taking $T$-duality into account, there are 15 different possibilities.

{}Next, we give the non-vanishing renormalizable terms in the tree-level
superpotential in the 
$M_{N_0,N_1,N_2,N_3}$ models
(we suppress the actual values of the Yukawa couplings, however):
\begin{eqnarray}
 {\cal W}=&& 
 \epsilon_{\alpha\beta\gamma} \Phi^s_\alpha X^s_\beta X^s_\gamma+
 \epsilon_{\alpha\beta\gamma} \phi^s_\alpha\chi^s_\beta {\widetilde \chi}^s_\gamma
 +y_{s s^\prime \alpha} \Phi^s_\alpha Q^{ss^\prime} Q^{ss^\prime} + 
 y_{s s^\prime \alpha} \phi^s_\alpha q^{ss^\prime} {\widetilde q}^{ss^\prime} +
 y_{ss^\prime \alpha} X^s_\alpha P^{ss^\prime} R^{ss^\prime}+ \nonumber\\
 && y_{ss^\prime \alpha} \chi^s_\alpha p^{ss^\prime} {\widetilde r}^{ss^\prime}+
 y_{ss^\prime \alpha} {\widetilde \chi}^s_\alpha {\widetilde p}^{ss^\prime}  
 {r}^{ss^\prime}+ P^{ss^\prime} {Q}^{s^\prime s^{\prime\prime}} R^{s^{\prime\prime} s}+
 p^{ss^\prime} {\widetilde q}^{s^\prime s^{\prime\prime}} r^{s^{\prime\prime} s}+
 {\widetilde p}^{ss^\prime} {q}^{s^\prime s^{\prime\prime}} 
 {\widetilde  r}^{s^{\prime\prime} s}~.
\end{eqnarray}
Here summation over the repeated indices is understood.
Also, $y_{ss^\prime\alpha}=\epsilon_{ss^\prime\alpha}$ if $s\not=0$, and 
$y_{0s^\prime\alpha}=\delta_{s^\prime\alpha}$. The condition $s\not=s^\prime\not=s^{\prime\prime}\not=s$ is implicitly assumed in the above expressions, and we are using the compact notation where $P^{s^\prime s}=P^{s s^\prime}$, 
$p^{s^\prime s}={\widetilde p}^{s s^\prime}$, $Q^{s^\prime s}=R^{s s^\prime}$, $q^{s^\prime s}=r^{s s^\prime}$,  and
${\widetilde q}^{s^\prime s}={\widetilde r}^{s s^\prime}$.
 
\subsection{Cases with Two and Three Wilson Lines}

{}In the previous subsection we constructed Type I vacua via compactification on the toroidal orbifold ${\cal M}=T^6/{\bf Z}_2\otimes{\bf Z}_2\otimes {\bf Z}_3$ with one Wilson line. The latter was chosen to be a ${\bf Z}_3$ valued shift in the $z_3$ direction. However, we can also consider adding Wilson lines acting as ${\bf Z}_3$ valued shifts in the $z_1$ and $z_2$ directions. Let us first consider the case where we have two Wilson lines. Let the first Wilson line be described as the shift $S$ in the $z_3$ direction as in the previous subsection. Let $U$ be a   
${\bf Z}_3$ valued shift in the $z_2$ direction. The action of $S$ on the Chan-Paton factors is given by matrices $\gamma_{S,p_s}$ as was described in the previous subsection. The action of $U$ on the Chan-Paton factors is similar to that of $S$ but must satisfy certain consistency conditions. Let the corresponding Chan-Paton matrices be $\gamma_{U,p_s}$. Note that $U$ commutes with $g$ and $R_2$, but generates non-Abelian groups isomorphic to $D_3$ together with $R_1$ and $R_2$. Also note that the following tadpole cancellation conditions must be satisfied ($k,k^\prime,k^{\prime\prime}=1,2$):   
\begin{equation}
 {\mbox{Tr}}(\gamma_{U^k  
 g^{k^\prime} , p_s})={\mbox{Tr}}(\gamma_{S^k U^{k^\prime}  
 g^{k^{\prime\prime}} , p_s})=-2~.
\end{equation}
Up to equivalent representations we have the following solutions.\\
$\bullet$ $N_s=0$ (for a given value of $s$). In this case the action of $S$ on the corresponding
D$p_s$-brane Chan-Paton factors is trivial. The action of $U$ on these Chan-Paton factors then is analogous to that of $S$ in the previous subsection and is given by: 
\begin{equation}
 \gamma_{U,p_s}={\mbox{diag}}({\bf X}\otimes {\bf I}_{2n_s}, 
 {\bf I}_{12-4n_s}, {\bf X} \otimes {\bf I}_{n_s}, {\bf I}_{4-2n_s})~,
\end{equation}
where ${\bf X}=-{1\over 2}{\bf I}_2 +{i\sqrt{3}\over 2}\sigma_2$, and $n_s$ can take values $n_s=0,1,2$. The gauge group coming from the D$p_s$-D$p_s$ open string sector (for this given value of $s$) is $[U(6-2n_s)\otimes U(n_s)^3\otimes Sp(4-2n_s)]_{ss}$. \\
$\bullet$ $N_s=1$ (for a given value of $s$). One possibility here is that $U$ acts trivially on the
corresponding Chan-Paton factors. In the case this action is non-trivial, we have: 
\begin{eqnarray}
 &&\gamma_{S,p_s}={\mbox{diag}}({\bf W}^\prime\otimes {\bf I}_{2}, 
 {\bf I}_{8}, {\bf W}^\prime , {\bf I}_{2})~,\\
 &&\gamma_{U,p_s}={\mbox{diag}}({\bf I}_4, {\bf X}\otimes {\bf I}_2, {\bf I}_{6},  {\bf X})~.
\end{eqnarray}
The gauge group coming from the D$p_s$-D$p_s$ open string sector (for this given value of $s$) is $[U(2)\otimes U(1)^6]_{ss}$.\\
$\bullet$ $N_s=2$ (for a given value of $s$). In this case the action of $U$ on the corresponding Chan-Paton factors must be trivial. The gauge group coming from the D$p_s$-D$p_s$ open string sector (for this given value of $s$) is $[U(2)^4]_{ss}$.

{}Next, we can consider adding the third Wilson line acting as a ${\bf Z}_3$ shift $V$ in the $z_1$ direction. From the above discussion it should be clear what the tadpole cancellation conditions are in this case (by ${\bf Z}_3$ symmetry between $S$, $U$ and $V$). In particular,
one of the tadpole cancellation conditions reads: 
\begin{equation}
 {\mbox{Tr}}(\gamma_{S U V g , p_s})=-2~.
\end{equation}
Up to equivalent representations we have the following solutions.\\
$\bullet$ $N_s=0$ and $n_s=0$ (for a given value of $s$). In this case the action of $S$ and $U$ on the corresponding
D$p_s$-brane Chan-Paton factors is trivial. The action of $V$ on these Chan-Paton factors then is given by: 
\begin{equation}
 \gamma_{V,p_s}={\mbox{diag}}({\bf Y}\otimes {\bf I}_{2m_s}, 
 {\bf I}_{12-4m_s}, {\bf Y} \otimes {\bf I}_{m_s}, {\bf I}_{4-2m_s})~,
\end{equation}
where ${\bf Y}=-{1\over 2}{\bf I}_2 +{i\sqrt{3}\over 2}\sigma_1$, and $m_s$ can take values $m_s=0,1,2$. The gauge group coming from the D$p_s$-D$p_s$ open string sector (for this given value of $s$) is $[U(6-2m_s)\otimes U(m_s)^3\otimes Sp(4-2m_s)]_{ss}$. \\
$\bullet$ $N_s=0,n_s=1$ or $N_s=1,n_s=0$ (for a given value of $s$). One possibility here is that $U$ acts trivially on the corresponding Chan-Paton factors. The case where this action is non-trivial is analogous to the case with two Wilson lines $S$ and $U$ with $N_s=1$ and $n_s=1$. The corresponding gauge group coming from the D$p_s$-D$p_s$ open string sector (for this given value of $s$) is then $[U(2)\otimes U(1)^6]_{ss}$.\\
$\bullet$ $N_s=n_s=1$ (for a given value of $s$). In this case the action of $V$ on the
corresponding
D$p_s$-brane Chan-Paton factors must be trivial. The gauge group coming from the D$p_s$-D$p_s$ open string sector (for this given value of $s$) is $[U(2)\otimes U(1)^6]_{ss}$.

{}It is completely straightforward to obtain the massless spectra and superpotentials of the above models with two and three Wilson lines starting from those with only one Wilson line by projecting onto the states invariant under the orbifold groups generated by $\gamma_{U , p_s}$ and $\gamma_{V , p_s}$. However, for the sake of brevity, here we will skip the details.

\section{Phenomenology of Three-Family Type I Vacua}

{}In the previous section we explicitly constructed various Type I vacua. To have a chance of being phenomenologically viable, these string models must contain the Standard Model gauge group $SU(3)_c\otimes SU(2)_w\otimes U(1)_Y$ as a subgroup. This constraint limits the number of models we need to study. For the sake of concreteness, in the following we will confine our attention to the models with one Wilson line only. The models with two and three Wilson lines appear to share the same generic features as those with one Wilson line, and can be analyzed in a similar fashion.    

{}For phenomenological reasons (which will become clear in a moment) it is advantageous to assume that the Standard Model gauge group comes from D5-branes. Without loss of generality we can assume that this gauge group (or at least the $SU(3)_c$ subgroup of it) comes from the D$5_1$-branes. Then for the D$5_1$-brane
gauge group to contain the $SU(3)_c$ subgroup we must assume that $N_1=0$ or $N_1=1$.
In the first case the $5_1 5_1$ gauge group is 
$U(6)\otimes Sp(4)$, whereas in the second case the $5_1 5_1$ gauge group is
$U(4)\otimes SU(2)\otimes U(1)^3$.

{}In the following we will first discuss some generic features of these models. Then we will discuss some possibilities for obtaining the Standard Model gauge group with three chiral generations from the corresponding compactifications. Finally, we will point out some
properties of these models which might be problematic phenomenologically. In particular, we 
will focus on the question of whether extra ``vector-like'' families can become heavy in these models. As we discuss below, this point is intimately related to the question of proton stability. 

\subsection{General Remarks} 

{}In Type I strings, the graviton (coming from the closed string sector) 
lives in the bulk, while the gauge and charged matter fields (coming 
from the open string sector)
live in the world-volumes of the branes.
Let $\lambda$ be the string
coupling, {\em i.e.}, $\lambda \sim e^{\phi}$, where
$\phi$ is the dilaton field.
At low energies, the 4-dimensional Planck mass $M_P$ and the
Newton's constant $G_N$ are given by
\begin{equation}\label{newton}
G_N^{-1}=M_P^2 =
{{8 M_S^8 v_1 v_2 v_3} \over \lambda^2}~,
\end{equation}
and the gauge couplings of the D9-brane and D$5_s$-brane fields are
\begin{equation}
g_9^{2} = {{(2 \pi)^4 \lambda} \over {M_S^6 v_1 v_2 v_3}}~, ~~~
g_s^{2} = {\lambda \over {M_S^2 v_s}}~,
\end{equation}
where $M_S$ is the string scale ({\em i.e.}, $M_S^{-2}$ is the string tension) and $v_s$ is the volume of the (orbifolded) two-torus parametrized by the complex coordinate $z_s$
($s=1,2,3$). These relations are subject to quantum and high energy corrections which 
we will ignore for the moment.

{}As can be seen from the above expressions, the gauge and gravitational couplings depend on the the dilaton and compactification volumes in different ways. This {\em a priori} allows them to unify at any scale between about $1~{\mbox{TeV}}$ and $10^{16}~{\mbox{GeV}}$ in Type I compactifications \cite{witten}. This observation, in particular, has opened up a possibility for
unification of all forces of nature at scales as low as $1~{\mbox{TeV}}$ \cite{TeV}. Scenarios of this type have dramatic experimental consequences which should be testable in the nearest future.

{}Unification of the gauge and gravitational couplings in Type I vacua ultimately requires that at least one of the volumes $v_s$ must be much larger than $M_S^{-2}$. We will thus assume that $v_3$ satisfies this requirement. Then it follows that the $99$ and $5_3 5_3$ gauge couplings are
weak. We  will need to assume that one of the volumes, which we choose to be $v_1$, is not that much different from $M_S^{-2}$ (so that the Standard Model gauge couplings are not too small).   

{}The main question we would like to address in the remainder of this section is whether it is possible to obtain the Standard Model gauge group with three chiral families from the Type I models discussed in the previous section. This discussion is going to be largely independent of the possible value of the string scale (at which all forces would have to unify). We will, however, point out in the appropriate cases the ``preferred'' values for the string scale. Before we go into more detail (and model dependent analyses), we would like to point out the key feature of these models, namely, how three-families of quarks and leptons can emerge from any of these Type I compactifications.

{}The reason why the above models can be thought of three-family models is the following. Suppose the $SU(3)_c$ subgroup of the Standard Model gauge group somehow occurs in the $5_1 5_1$ sector. Then it is not difficult to see that the fields $X^1_\alpha$ in Table I contain three copies of fields transforming in the irrep $({\bf 3},{\bf 2})$ of $SU(3)_c\otimes SU(2)_w$. (In fact, in the scenarios we discuss below $SU(2)_w$ also arises as a subgroup of the $5_1 5_1$ gauge group.) This ultimately implies that there are three chiral families in such a model. This follows from the requirement of non-Abelian gauge anomaly cancellation. In particular, regardless of the details of a given model, there must remain six fields transforming in $({\overline {\bf 3}},{\bf 1})$ of $SU(3)_c\otimes SU(2)_w$. Nonetheless, this is not sufficient to conclude that we indeed have the Standard Model gauge group with three families. First, it is far from being obvious that the $U(1)_Y$ exists, and if it does, then whether the hypercharge assignments are correct. Second, it is not guaranteed that there are no extra ``vector-like'' families of, say, fields transforming in the irreps $({\bf 3},{\bf 1})$ and $({\overline {\bf 3}},{\bf 1})$. There are no such extra fields in the Standard Model, so for the above models to be realistic one would need to
get rid of them one way or another. Finally, we need electroweak Higgs doublets to be present in the model. All of these requirements need to be satisfied for the model to be realistic. In the following we will investigate some of these issues.       

\subsection{The $U(4)\otimes SU(2)\otimes U(1)^3$ Models}

{}Let us first consider the models $M_{N_0, 1, N_2, N_3}$. In these models the D$5_1$-brane gauge group is $U(4)\otimes SU(2)\otimes U(1)^3$. For definiteness we will focus on the model
$M_{2,1,0,2}$ with the gauge group $[U(2)^4]_{99} \otimes [U(4)\otimes SU(2)\otimes U(1)^3]_{5_1 5_1}\otimes [U(6)\otimes Sp(4)]_{5_2 5_2}\otimes [U(2)^4]_{5_3 5_3}$. The reason why we would like to discuss this particular model here is that we would like to point out a possible scenario for obtaining three chiral families of quarks and leptons in this model. This scenario is specific to the fact that the gauge group of one of the D5-branes is $U(6)\otimes Sp(4)$. Other models of the type $M_{N_0, 1, N_2, N_3}$ can be analyzed similarly (albeit different models have their own peculiarities). In the subsequent sections, however, we will point out some of the generic features that all of these (as well as other) models appear to share. The discussion in this subsection, therefore, should be viewed as an attempt to understand what kind of possibilities these models offer as far as embedding of three chiral families of the Standard Model is concerned.  

{}As pointed out earlier, $g_9$ of the gauge group $[U(2)^4]_{99}$ and $g_3$ of the the gauge group $[U(2)^4]_{5_3 5_3}$ are expected to be very weak. Furthermore, there are enough fields to Higgs these gauge groups completely. So we 
will ignore them to simplify the discussions. Note that the first $U(1)$ in the D$5_1$ gauge group is anomalous. This $U(1)$ will therefore be broken\footnote{If the string scale is of order $1~{\mbox{TeV}}$, the ``naive'' anomalous $U(1)$ breaking scale may turn out to be around (or even lower than) the supersymmetry breaking scale. In this case it is not completely clear that the anomalous $U(1)$ is indeed broken since cancellation of the corresponding D-term may not be favorable energetically due to relatively large soft masses for the corresponding fields.} by the generalized Green-Schwarz mechanism. Furthermore, note that the other three $U(1)$'s are irrelevant as far as the Standard Model gauge group is concerned. In particular, the fields $X^1_\alpha = ({\bf 4}, {\bf 2})$ (which are supposed to give rise to three families of left-handed quarks and lepton doublets) are not charged under these three $U(1)$'s. In fact, these $U(1)$'s can be broken by giving appropriate vevs to the fields
$\phi^1_\alpha,\chi^1_\alpha,{\widetilde \chi}^1_\alpha$. Thus, in the $5_1 5_1$ sector we are left with the gauge group $SU(4)\otimes SU(2)$. In fact, we will identify $SU(4)$ with $SU(4)_c
(\supset SU(3)_c)$, and $SU(2)$ with $SU(2)_w = SU(2)_L$.    

{}As we have already mentioned above, there are three chiral families of  
$({\bf 4}, {\bf 2})$ of $SU(4) \otimes SU(2)$. To have three chiral families of quarks and leptons, we also must find the corresponding six copies of fields transforming in
$({\overline {\bf 4}}, {\bf 1})$ of $SU(4) \otimes SU(2)$. These are obviously not present in the $5_1 5_1$ sector. We will look for them in the $5_1 5_2$ sector. In particular, there are 
$5_1 5_2$ sector matter fields transforming in the irrep $({\overline {\bf 4}}_1,{\bf 1}_1; 
{\overline {\bf 6}}_2,{\bf 1}_2)$ of the gauge group $[SU(4) \otimes SU(2)]_{5_1 5_1}\otimes [SU(6) \otimes Sp(4)]_{5_2 5_2}$. After appropriately Higgsing the ${5_2 5_2}$ gauge group\footnote{Note that the $U(1)$ subgroup of the original $U(6)\otimes Sp(4)$ gauge group in the $5_2 5_2$ sector is anomalous an is automatically broken.}, we can attempt to identify these with the required fields of the Standard Model as discussed above.  
 
{}The required Higgsing in the ${5_2 5_2}$ sector is the following. 
First, let us consider the breaking
$SU(6) \supset Sp(6) \supset SU(2) \otimes Sp(4) \supset SU(2) \otimes 
SU(2) \otimes SU(2) \supset SU(2)$. This is achieved by giving 
appropriate vevs to the three ${\overline {\bf 15}}$'s of $SU(6)$ : 
\begin{equation}
{\overline {\bf 15}} = {\bf 1} \oplus {\bf 14} \\
 = 2({\bf 1},{\bf 1}) \oplus ({\bf 1},{\bf 5}) \oplus ({\bf 2},{\bf 4})\\
 = 3({\bf 1},{\bf 1},{\bf 1}) \oplus ({\bf 1},{\bf 2},{\bf 2}) \oplus
	 ({\bf 2},{\bf 2},{\bf 1}) \oplus ({\bf 2},{\bf 1},{\bf 2})~. 
\end{equation}
There are three ${\overline {\bf 15}}$'s of $SU(6)$, so we can use one to break 
$SU(6) \supset Sp(6)$, one to break $Sp(6) \supset SU(2) \otimes Sp(4)$,
and the remaining one to break $SU(2) \otimes Sp(4) \supset SU(2) \otimes
SU(2) \otimes SU(2)$. This mechanism leaves the following set of charged matter fields
behind: $({\bf 1},{\bf 2},{\bf 2})$, $({\bf 2},{\bf 2},{\bf 1})$ and 
$({\bf 2},{\bf 1},{\bf 2})$, plus singlets. The bi-fundamentals of $SU(2)^3$ can acquire vevs to break it to $SU(2)_{\mbox{\small diag}}$ (with one adjoint of this 
$SU(2)_{\mbox{\small diag}}$ remaining), or even $U(1)$. In the first case we can identify $SU(2)_{\mbox{\small diag}}$ with $SU(2)_R$, and treat the gauge subgroup $SU(4)_c\otimes
SU(2)_w\otimes SU(2)_R$ (where the first two factors come from the $5_1 5_1$ sector, whereas the third one comes from the $5_2 5_2$ sector) as the Pati-Salam gauge group that contains the Standard Model gauge group as a subgroup. In the second case the corresponding gauge group 
is $SU(4)_c\otimes SU(2)_w\otimes U(1)$ which also contains the Standard Model gauge group.
In this scenario we will need to assume that the volume $v_2$ is not too much different from $v_1$.

{}Let us see what happens to the $5_1 5_2$ matter field 
$(\overline{\bf 4}_1,{\bf 1}_1;\overline{\bf 6}_2,{\bf 1}_2)$ (where the irreps correspond to
$[SU(4)\otimes SU(2)]_{5_1 5_1}\otimes [SU(6)\otimes Sp(4)]_{5_2 5_2}$). For concreteness let us concentrate on the case where $SU(6)$ is broken to $SU(2)_R$ in the above Higgsing process. (The case where $SU(2)_R$ is further broken to $U(1)$ is similar.)
Under $SU(6) \supset SU(2) \otimes SU(2) \otimes SU(2)$, it becomes
$(\overline{\bf 4},{\bf 2},{\bf 1},{\bf 1}) \oplus
(\overline{\bf 4},{\bf 1},{\bf 2},{\bf 1}) \oplus
(\overline{\bf 4},{\bf 1},{\bf 1},{\bf 2})$.
Under $SU(2) \otimes SU(2) \otimes SU(2) \supset SU(2)_R$, they become
three copies of $(\overline{\bf 4},{\bf 2})$ of $SU(4) \otimes SU(2)_R$
inside the Pati-Salam gauge group, that is, they become
three copies of $(\overline{\bf 4}_1,{\bf 1}_1;{\bf 2}_2,{\bf 1}_2)$ of $[SU(4) \otimes SU(2)]_{5_1 5_1}\otimes [SU(2)\otimes Sp(4)]_{5_2 5_2}$. These, together with the fields
$X^1_\alpha$, give three families of quarks and leptons transforming in the appropriate irreps of the Pati-Salam gauge group.

{}Here the following remark is in order. It is not difficult to see that (before the above Higgsing) the field $R^{12}$ is a potential candidate for producing the electroweak doublets. Via a careful examination of the superpotential given in the previous section it is not difficult to see that this field acquires mass once the above Higgsing $SU(6)\supset SU(2)_R (\supset U(1))$ is performed in the $5_2 5_2$ sector. 
(For concreteness let us concentrate on the case where we have the Pati-Salam gauge group.)
As a result we get 3 copies of massive fields in the irrep
$({\bf 1}_1,{\bf 2}_1;{\bf 2}_2,{\bf 1}_2)$ of $[SU(4) \otimes SU(2)]_{5_1 5_1}\otimes [SU(2)\otimes Sp(4)]_{5_2 5_2}$. These fields could {\em a priori} give rise to the electroweak Higgs doublets (which would have to radiatively acquire negative mass squared once supersymmetry is broken). 

{}The above features of this model are compatible with phenomenology. However, there are certain shortcomings in this model, to which we now turn. Note that the only other fields that are charged under the Pati-Salam gauge group are the following: $P^{01}$, $P^{13}$, $Q^{01}$,
 $Q^{12}$, $Q^{13}$, $R^{13}$. None of these fields, however, have the correct quantum numbers to break the Pati-Salam gauge group to that of the Standard Model.  This is obvious for the fields carrying superscripts other than 1 and 2: such fields are charged under $99$ and $5_3 5_3$ gauge group. (The latter are very weakly coupled and can be considered to be ``global'' symmetries. This however does not change the state of affairs as will become clear in a moment.) On the other hand, the field $Q^{12}$ is charged under the $Sp(4)$ subgroup of the
$5_2 5_2$ gauge group. This field, therefore, cannot be used in the breaking of the Pati-Salam gauge group to that of the Standard Model either. To achieve such a breaking we would need a ``vector-like'' field transforming in the irreps 
$({\bf 4}_1,{\bf 1}_1;{\bf 2}_2,{\bf 1}_2)$ and 
$(\overline{\bf 4}_1,{\bf 1}_1;{\bf 2}_2,{\bf 1}_2)$ of 
$[SU(4) \otimes SU(2)]_{5_1 5_1}\otimes [SU(2)\otimes Sp(4)]_{5_2 5_2}$. Such fields are simply absent in the spectrum of the above model. Thus, it may be problematic to break the Pati-Salam 
gauge group to $SU(3)_c \otimes SU(2)_w \otimes U(1)_Y$ in the above 
scenario\footnote{Here we would like to point out that {\em a priori} there might be a way around this difficulty. Thus, if the string scale $M_S$ is in the TeV range, then one could use the scalar superpartners of chiral fermions in $({\overline{\bf 4}}_1,{\bf 1}_1;{\bf 2}_2,{\bf 1}_2)$ to break the Pati-Salam gauge group down to that of the Standard Model. In particular, this need not be in conflict with F- and D-flatness conditions if supersymmetry is also broken around the same scale.}. This is somewhat similar to the situation (which was pointed out in \cite{4-2-1}) in the Pati-Salam type of model of \cite{ST,class} in the scenario which leads to three chiral families \cite{ST}. 

\subsection{The $U(6)\otimes Sp(4)$ Models}

{}The problem we pointed out in the previous subsection can be avoided in another set of models, namely, in the $M_{N_0,0,N_1,N_2}$ models, where the $5_1 5_1$ gauge group is $U(6)\otimes Sp(4)$. (One of the models of this type, namely, the $M_{0,0,0,0}$ model, was originally constructed in \cite{su6}.) There are three families in the $SU(6)$ subgroup in these models. However, as was pointed out in \cite{su6}, the gauge subgroup $SU(6)$ cannot really be treated in these models as a ``grand unified'' gauge group for the reason that there are no appropriate matter fields in these models to break $SU(6)$ to the Standard Model gauge group. 
There is, however, another possibility here to which we now turn. (This possibility was pointed out in \cite{ST} in the context of the model of \cite{su6}.)

{}We need not assume that the Standard Model gauge group comes solely from the $SU(6)$ subgroup. Instead, we can attempt to embed the former in $SU(6)\otimes Sp(4)$. (In the following we are going to ignore the anomalous $U(1)$ subgroup.) This can be achieved as follows.
Consider the branchings of $({\overline {\bf 15}},{\bf 1})$ and $({{\bf 6}},{\bf 4})$ of 
$SU(6)\otimes Sp(4)$ under the breaking $SU(6)\otimes Sp(4)\supset [SU(4)\otimes SU(2)\otimes U(1)]\otimes [SU(2)\otimes SU(2)]$ (the $U(1)$ charges are given in parentheses): 
\begin{eqnarray}
 &&({\overline {\bf 15}},{\bf 1}) = ({\bf 6},{\bf 1},{\bf 1},{\bf 1})(-2)\oplus
 ({\bf 1},{\bf 1},{\bf 1},{\bf 1})(+4)\oplus ({\overline {\bf 4}},{\bf 2},{\bf 1},{\bf 1})(+1)~,\\
 &&({{\bf 6}},{\bf 4}) = ({\bf 4},{\bf 1},{\bf 2},{\bf 1})(+1)\oplus
 ({\bf 4},{\bf 1},{\bf 1},{\bf 2})(+1)\oplus
 ({\bf 1},{\bf 2},{\bf 2},{\bf 1})(-2)\oplus ({\bf 1},{\bf 2},{\bf 1},{\bf 2})(-2)~.  
\end{eqnarray}
Now consider giving vevs to $({\bf 1},{\bf 1},{\bf 1},{\bf 1})(+4)$ in $\Phi^1_1$ and $({\bf 1},{\bf 2},{\bf 2},{\bf 1})(-2)$ in $X^1_1$. 
It is not difficult to see that the unbroken gauge symmetry is then
$SU(4)\otimes SU(2)\otimes SU(2)$ which we identify with the Pati-Salam gauge group.
Determining the matter content requires some care as there are non-trivial couplings in the superpotential. The resulting massless matter is given by 
 \begin{eqnarray}
 3\times({\bf 6},{\bf 1},{\bf 1})\oplus
 3\times ({\bf 4},{\bf 2},{\bf 1})\oplus 
 1\times ({{\bf 4}},{\bf 1},{\bf 2})\oplus
 1\times ({\overline {\bf 4}},{\bf 1},{\bf 2})\oplus 1\times ({{\bf 1}},{\bf 2},{\bf 2})~,
\end{eqnarray}
plus singlets. The irreps are of the resulting $SU(4)_c\otimes SU(2)_w \otimes SU(2)_R$ Pati-Salam gauge group which comes solely from the D$5_1$-branes. A nice feature of the above spectrum is that it contains the ``vector-like'' matter $({{\bf 4}},{\bf 1},{\bf 2})\oplus  ({\overline {\bf 4}},{\bf 1},{\bf 2})$ which is exactly what is required to break the Pati-Salam gauge group to $SU(3)_c\otimes SU(2)_w \otimes U(1)_Y$. Moreover, there is matter in
$({{\bf 1}},{\bf 2},{\bf 2})$ which would give rise to the electroweak Higgs doublets.

{}Although the above features make this model look attractive, we would like to point out the following setback. Unlike in the models of the previous subsection, here we have no obvious (or natural) way of identifying what is required to complete the three chiral generations of quarks and leptons. The missing fields should transform in (three copies of) $({\overline {\bf 4}},{\bf 1},{\bf 2})$ of $SU(4)_c\otimes SU(2)_w \otimes SU(2)_R$. These fields are not found in the $5_1 5_1$ sector, and for obvious reasons they cannot come from $5_1 9$, $5_1 5_2$ or $5_1 5_3$ sectors. Thus, compared with the $U(4)\otimes SU(2)\otimes U(1)^3$ models of the previous subsection, the $U(6)\otimes Sp(4)$ models have the advantage that one can break the gauge group down to that of the Standard Model. However, the generation structure does not appear to
give rise to three families of quarks and leptons. In fact, in the next subsection we will point out a more generic feature common to all of the above models which is related to the problems with the generation structure discussed in this subsection.

\subsection{``Vector-Like'' Matter and Baryon Number Violation}

{}Regardless of a particular scenario, the above models cannot be realistic unless there are
no light chiral fermions transforming in the irrep $({\bf 3},{\bf 1})$ of $SU(3)_c\otimes SU(2)_w$ (as the latter type of states are absent in the Standard Model). However, in all of these models we have a potential source for such states coming from ${\bf 6}$'s of $SU(4)_c$. 
It is not clear how to make these states massive. However, instead of discussing the problem with the triplets coming from ${\bf 6}$'s of $SU(4)_c$, we will focus on the antitriplets arising from
these ${\bf 6}$'s. These states are dangerous for the following reasons. From the superpotential, namely, from the couplings $\epsilon_{\alpha\beta\gamma} \Phi^s_\alpha X^s_\beta X^s_\gamma$, it is not difficult to see that they will give rise to baryon number violating dimension four
operators. (More concretely, these couplings have the form $({\bf 3},{\bf 1})\cdot ({\bf 1},{\bf 2})
\cdot({\overline {\bf 3}},{\bf 2})$.) It is, therefore necessary to make sure that these states acquire
large enough masses.

{}It is not difficult to see that the only way\footnote{We will discuss another non-trivial possibility in a moment.}  we can get rid of these states is via the couplings of the type $y_{s s^\prime \alpha} \Phi^s_\alpha Q^{ss^\prime} Q^{ss^\prime}$ in the superpotential. This coupling, however, implies that to make {\em all three} of the antitriplets coming from $\Phi^1_\alpha$ (recall that by our convention $SU(3)_c$ comes from the D$5_1$-branes) heavy, it is required that $R^{01}$, $Q^{12}$ and $Q^{13}$ acquire vevs. Moreover, since these states do not exist for $N_0=N_2=N_3=2$, we are bound to consider only the models with $N_0,N_1,N_2,N_3=0,1$.
We cannot allow any mixing between the $5_1 5_1$ and other D-brane gauge groups (as this will destroy the three-family feature of the model). On the other hand, in order for 
$R^{01}$, $Q^{12}$ and $Q^{13}$ to carry no quantum numbers (except, perhaps, for $U(1)$ charges) under these other D-brane gauge groups, we must (in the models with $N_0,N_1,N_2,N_3=0,1$) Higgs the 99, $5_2 5_2$ and $5_3 5_3$ gauge groups completely\footnote{Actually, since the 99 and $5_3 5_3$ gauge groups are very weakly coupled, it might not be necessary to Higgs them. However, it is definitely required to Higgs the $5_2 5_2$ gauge group.}. There is no obstruction to Higgsing these gauge groups in any of these models. However, at the end of the day we are left at best with the Standard Model gauge group but no other non-Abelian gauge subgroups. It is then unclear what would be the source of supersymmetry breaking in these models. 

{}Note that even if we Higgs these extra gauge groups completely, it is still not guaranteed that we will obtain the desired generation structure or get rid of all the other unwanted states. Thus, even if we get rid of the antitriplets, as we already mentioned, it is far from being obvious how to get rid of their triplet counterparts (also coming from ${\bf 6}$'s of $SU(4)_c$). These states would be disastrous if they remain light. 

{}Before we finish our discussion here, we would like to comment on a possibility of making ${\bf 6}$'s of $SU(4)_c$ heavy via couplings involving not only the open string sector states but also those coming from the closed string sectors. Thus, ${\bf 6}\cdot{\bf 6}$ contains a singlet of $SU(4)_c$, so naively it might seem that there may be operators of the type ${\bf 6}\cdot{\bf 6} \cdot \Sigma$, where $\Sigma$ are some operators in the closed string sector. Then if $\Sigma$ has a vev, ${\bf 6}$'s become heavy. It, however, should be clear that this cannot be the case simply from the fact that we are considering perturbative Type I compactifications, and in such backgrounds we do not expect ``mass terms'' of this type on very general grounds. Nonetheless, the issue involved here appears to be non-trivial, so we would like to elaborate this issue in a bit more detail.

{}The point is that some of the twisted closed string states transform non-trivially under the anomalous $U(1)$ in the $5_1 5_1$ sector. On the other hand, ${\bf 6}\cdot{\bf 6}$ operator carries non-zero charge under this $U(1)$. To construct a gauge invariant operator (which could potentially give rise to a mass term) we need to compensate for this $U(1)$ charge. There are ${\bf Z}_3$ twisted closed string singlets which upon certain non-linear transformation (which is related to the chiral-linear multiplet duality transformation in four dimensions) can be written in a basis where they carry non-zero charges under the anomalous $U(1)$. This can be most conveniently checked starting from Type I on the ${\bf Z}_3$ orbifold \cite{Sagnotti}, and constructing the heterotic dual of this model as in \cite{Sagnotti,ZK}. Then the 27 ${\bf Z}_3$ twisted closed string singlets in the heterotic picture are explicitly charged under the anomalous $U(1)$ subgroup of the 99 Chan-Paton gauge group $U(12)\otimes SO(8)$. The 
couplings in this model relevant for our discussion here read (we are being schematic here)  $({\overline {\bf 66}},{\bf 1})^6\cdot \Theta^3$, where $\Theta$ are the 27 ${\bf Z}_3$ twisted closed string singlets. Upon adding the ${\bf Z}_2\otimes {\bf Z}_2$ twists and the Wilson line, we end up with the coupling of the type $(\Phi^2 \phi)^2 \Theta^3$ (here the number of surviving singlets $\Theta$ after the appropriate projections is 10), so that there is an effective quartic coupling between $\Phi$'s (once $\Theta$'s and $\phi$'s get vevs), but there is no mass term for $\Phi$'s.
One simple way to see this result is to note that there is a ${\bf Z}_2$ discrete symmetry under which the fields $\Phi$ and $\phi$ have charge $-1$, whereas $\Theta$'s have charge $+1$.
This follows from the non-Abelian embedding of the ${\bf Z}_2\otimes {\bf Z}_2$ projection into the Chan-Paton factors.

\section{Conclusions}

{}In the previous section we have seen that there are many non-trivial 
issues arising in the
Type I models classified in this paper. Thus, in some cases it is difficult to break the
gauge group down to $SU(3)_c\otimes SU(2)_w \otimes U(1)_Y$. In other cases this does not seem to be a problem, but the three families one obtains are not necessarily coincident with three generations of quarks and leptons of the Standard Model. One of the problems that appears to be quite generic is the presence of extra light states which give rise to baryon number violating couplings. In fact, this problem seems to be quite generic for even a larger class of Type I vacua. In particular, it was recently encountered in \cite{lykken} where a three-family $SU(5)$ ``GUT'' was constructed from the orientifold of Type IIB on the ${\bf Z}_3$ orbifold with D3-branes. This model is not realistic as it is not possible to break $SU(5)$ down to the Standard Model gauge group. One interesting feature of this model is that it too suffers from the baryon number violating terms in a fashion very similar to the models we discussed in this paper.

{}One of the open question as far as phenomenology of perturbative Type I vacua is concerned is that of supersymmetry breaking. It is especially non-trivial in the context of the TeV scale string unification. The traditional gravity mediated supersymmetry breaking scenario does not seem to be adequate here. Gauge mediated supersymmetry breaking might be a possibility, but so far it is unclear how it would be realized. Not knowing much about supersymmetry breaking mechanisms (if any) in such models, it is rather difficult to definitively study their phenomenological implications. In particular, some of the observations we made in this paper implicitly rely on certain (perhaps, reasonable) assumptions about the dynamics. However, at present it is not clear how robust such conclusions can be, especially in the context of TeV string unification where supersymmetry breaking and possibly spontaneous breaking of (larger) gauge symmetry, as well as anomalous $U(1)$ effects can be intertwined in a non-trivial fashion. In particular, some of the intuition valid in the standard GUT scenario might not be directly applicable in these cases.

{}It is clear that a better understanding of supersymmetry breaking in Type I compactifications is more than desirable. It is also conceivable that we can start from Type I models without supersymmetry (which {\em a priori} is reasonable in the context of TeV string unification). It might therefore be important to understand non-supersymmetric chiral Type I vacua in four dimensions, and perhaps construct some explicit examples of such models.      

\acknowledgements

{}We would like to thank Pran Nath and Gary Shiu for discussions.
The research of S.-H.H.T. was partially
supported by the National Science Foundation.
The work of Z.K. was supported in part by the grant NSF PHY-96-02074,
and the DOE 1994 OJI award.
Z.K. would also like to thank Albert and Ribena Yu for
financial support.

\begin{table}[t]
\begin{tabular}{|c|c|l|}
 Model and Gauge Group& Field & Charged Matter  
  \\
 \hline
    &&\\
 ${\bf Z}_2\otimes{\bf Z}_2\otimes {\bf Z}_3$ &$\Phi^s_\alpha$ &
 $3\times [({\overline{\bf  A}}_s,{\bf 1}_s,{\bf 1}_s,{\bf 1}_s,{\bf 1}_s)_L]_{ss}$
 \\
 &$\phi^s_\alpha$ &
 $3\times [({{\bf  1}}_s,{\overline{\bf N}}_s,{\overline{\bf N}}_s,{\bf 1}_s,{\bf 1}_s)_L]_{ss}$  \\
 $\bigotimes_{s=0}^3 [U(M_s)\otimes U(N_s)^3\otimes Sp(L_s)]_{ss}$ &$X^s_\alpha$ &
 $3\times [({{\bf  M}}_s,{{\bf 1}}_s,{{\bf 1}}_s,{\bf 1}_s,{\bf L}_s)_L]_{ss}$  \\
  &$\chi^s_\alpha$ &
 $3\times [({{\bf  1}}_s,{{\bf N}}_s,{\bf 1}_s,{\overline{\bf N}}_s,{\bf 1}_s)_L]_{ss}$  \\ 
 &${\widetilde \chi}^s_\alpha$ &
 $3\times [({{\bf  1}}_s,{\bf 1}_s,{{\bf N}}_s,{{\bf N}}_s,{\bf 1}_s)_L]_{ss}$  \\ 
&$P^{s s^\prime}$ &
 $[({\overline{\bf  M}}_s,{\bf 1}_s,{\bf 1}_s,{\bf 1}_s,{\bf 1}_s;
  {\overline{\bf  M}}_{s^\prime},{\bf 1}_{s^\prime},{\bf 1}_{s^\prime},{\bf 1}_{s^\prime},{\bf  
 1}_{s^\prime})_L]_{ss^\prime}$  \\
 &$p^{s s^\prime}$ &
 $[({\bf 1}_s,{\overline{\bf  N}}_s,{\bf 1}_s,{\bf 1}_s,{\bf 1}_s;
 {\bf 1}_{s^\prime},{\bf 1}_{s^\prime},{\overline{\bf  N}}_{s^\prime},{\bf 1}_{s^\prime},{\bf 1}_{s^\prime})_L]_{ss^\prime}$  \\
 &${\widetilde p}^{ss^\prime}$ &
 $[({\bf 1}_s,{\bf 1}_s,{\overline{\bf  N}}_s,{\bf 1}_s,{\bf 1}_s;
 {\bf 1}_{s^\prime},{\overline{\bf  N}}_{s^\prime},{\bf 1}_{s^\prime},{\bf 1}_{s^\prime},{\bf 1}_{s^\prime})_L]_{ss^\prime}$  \\
  &$Q^{s s^\prime}$ &
 $[({{\bf  M}}_s,{\bf 1}_s,{\bf 1}_s,{\bf 1}_s,{\bf 1}_s;
 {\bf 1}_{s^\prime},{{\bf  1}}_{s^\prime},{\bf 1}_{s^\prime},{\bf 1}_{s^\prime},{\bf L}_{s^\prime})_L]_{ss^\prime}$\\
   &$q^{s s^\prime}$ &
 $[({\bf 1}_s,{{\bf  N}}_s,{\bf 1}_s,{\bf 1}_s,{\bf 1}_s;
 {\bf 1}_{s^\prime},{\bf 1}_{s^\prime},{\bf 1}_{s^\prime},{\overline{\bf  N}}_{s^\prime},{\bf 1}_{s^\prime})_L]_{ss^\prime}$  \\
  &${\widetilde q}^{s s^\prime}$ &
 $[({\bf 1}_s,{\bf 1}_s,{{\bf  N}}_s,{\bf 1}_s,{\bf 1}_s;
 {\bf 1}_{s^\prime},{\bf 1}_{s^\prime},{\bf 1}_{s^\prime},{{\bf  N}}_{s^\prime},{\bf 1}_{s^\prime})_L]_{ss^\prime}$  \\
   &$R^{s s^\prime}$ &
 $[({\bf 1}_s,{\bf 1}_s,{{\bf  1}}_s,{\bf 1}_s,{\bf L}_s;
 {\bf M}_{s^\prime},{\bf 1}_{s^\prime},{\bf 1}_{s^\prime},{{\bf  1}}_{s^\prime},{\bf 1}_{s^\prime})_L]_{ss^\prime}$  \\
   & $r^{s s^\prime}$ &
 $[({\bf 1}_s,{\bf 1}_s,{{\bf  1}}_s,{\overline {\bf N}}_s,{\bf 1}_s;
 {\bf 1}_{s^\prime},{\bf N}_{s^\prime},{\bf 1}_{s^\prime},{{\bf  1}}_{s^\prime},{\bf 1}_{s^\prime})_L]_{ss^\prime}$  \\ 
    & ${\widetilde r}^{s s^\prime}$ &
 $[({\bf 1}_s,{\bf 1}_s,{{\bf  1}}_s,{{\bf N}}_s,{\bf 1}_s;
 {\bf 1}_{s^\prime},{\bf 1}_{s^\prime},{\bf N}_{s^\prime},{{\bf  1}}_{s^\prime},{\bf 1}_{s^\prime})_L]_{ss^\prime}$  \\\hline
\end{tabular}
\caption{The massless open string spectra of the ${\cal N}=1$ Type I compactifications on $T^6/{\bf Z}_2\otimes{\bf Z}_2\otimes {\bf Z}_3$ with one Wilson line. Here $s=0,1,2,3$ labels the D$p_s$-branes, where $p_0=9$, and $p_s=5_s$ for $s=1,2,3$. The subscript ``$ss$''
indicates that the corresponding state comes from the D$p_s$-D$p_s$ open string sector.
Similarly, the subscript ``$ss^\prime$'' corresponds to the D$p_s$-D$p_{s^\prime}$ open string states (with $s<s^\prime$). Also, $\alpha=1,2,3$ labels the multiplicity of states in the 
D$p_s$-D$p_s$ sectors.
The notation ${\bf A}_s$ stands for the two-index ($M_s(M_s-1)/2$ dimensional) antisymmetric representation of $U(M_s)$. The corresponding $U(1)$ charge of the states ($\Phi^s_\alpha$) transforming in the ${\overline {\bf A}}_s$ irrep
is $-2$. Each fundamental ${\bf K}$ of $U(K)$ ($K=M_s,N_s$) is accompanied by the corresponding $U(1)$ charge $+1$, whereas the same $U(1)$ charge for the anti-fundamental ${\overline {\bf K}}$ is $-1$.
We are using the convention where $Sp(L_s)$ has rank $L_s/2$. The allowed values of
$M_s$, $N_s$ and $L_s$ (which depend on the choice of the Wilson line) are given by:
$M_s=6-2N_s$, $L_s=4-2N_s$, and $N_s=0,1,2$.
For $N_s=0$, $U(N_s)^3$ as well as the states with factors ${\bf N}_s$ or ${\overline {\bf N}}_s$ are absent. For $N_s=1$, ${\bf N}_s$ (${\overline {\bf N}}_s$) signifies the corresponding  $U(1)$ charge $+1$ ($-1$).}
\label{Z223} 
\end{table}

\end{document}